\documentclass{iau}

\usepackage{amsmath}
\usepackage{graphicx}
\usepackage{multirow}
\usepackage{natbib}

\newcommand\aap{A\&A}                
\newcommand\aapr{A\&ARv}             
\newcommand\aj{AJ}                   
\newcommand\apj{ApJ}                 
\newcommand\apjl{ApJ}                
\newcommand\apss{Ap\&SS}             
\newcommand\mnras{MNRAS}             
\newcommand\nat{Nature}              
\newcommand\prl{Phys. Rev.~Lett.}    
\newcommand\pasa{Publ. Astron. Soc. Australia}  
\newcommand\pasp{PASP}               
\newcommand\rmxaa{Rev. Mex. Astron. Astrofis.} 

\begin{document}

\lefttitle{J. E. M\'endez-Delgado \&  J. Garc\'ia-Rojas}
\righttitle{The abundance discrepancy in ionized nebulae: which are the correct abundances?}

\journaltitle{Planetary Nebulae: a Universal Toolbox in the Era of Precision Astrophysics}
\jnlDoiYr{2023}
\doival{10.1017/xxxxx}
\volno{384}

\aopheadtitle{Proceedings IAU Symposium}
\editors{O. De Marco, A. Zijlstra, R. Szczerba, eds.}

\title{The abundance discrepancy in ionized nebulae: which are the correct abundances?}

\author{J. E. M\'endez-Delgado$^{1}$ \& J. Garc\'ia-Rojas$^{2,3}$}
\affiliation{$^1$Astronomisches Rechen-Institut, Zentrum f\"ur Astronomie der Universit\"at Heidelberg, Mönchhofstraße 12-14, D-69120 Heidelberg, Germany\\ $^2$Instituto de Astrof\'isica de Canarias, E-38205 La Laguna, Tenerife, Spain\\ $^3$Departamento de Astrof\'isica, Universidad de La Laguna, E-38206 La Laguna, Tenerife, Spain}

\begin{abstract}

Ionized nebulae are key to understanding the chemical composition and evolution of the Universe. Among these nebulae, H~{\sc ii} regions and planetary nebulae are particularly important as they provide insights into the present and past chemical composition of the interstellar medium, along with the nucleosynthetic processes involved in the chemical evolution of the gas. However, the heavy-element abundances derived from collisional excited lines (CELs) and recombination lines (RLs) do not align. This longstanding abundance-discrepancy problem calls into question our absolute abundance determinations. Which of the lines (if any) provides the correct heavy-element abundances?
Recently, it has been shown that there are temperature inhomogeneities concentrated within the highly ionized gas of the H~{\sc ii} regions, causing the reported discrepancy. However, planetary nebulae do not exhibit the same trends as the H~{\sc ii} regions, suggesting a different origin for the abundance discrepancy. In this proceedings, we briefly discuss the state-of-the-art of the abundance discrepancy problem in both H~{\sc ii} regions and planetary nebulae.

\end{abstract}

\begin{keywords}
ISM: abundances, (ISM:) HII regions, (ISM:) planetary nebulae: general
\end{keywords}

\maketitle

\section{Introduction}

At the beggining of the universe, baryonic matter was composed exclusively of hydrogen, helium and small amounts of lithium \citep{Peebles:1966, Peebles:1966b, Wagoner:1973}. The rest of the heavier elements were formed later through nucleosynthetic processes related to the life and death of stars \citep{Bethe:1939, Caughlan:1965, Truran:1971}. Due to this interconnection between star formation, nucleosynthesis, and chemical composition, the study of the chemical abundances of ionized nebulae in the interstellar medium (ISM) has been crucial in understanding galaxies formation and evolution \citep{Peimbert:2017, Maiolino:2019}. For this purpose, star-forming regions, known as H~{\sc ii} regions, and planetary nebulae (PNe) are particularly important. The former are clouds of gas ionized by the ultraviolet radiation emitted by newly formed massive O type or early B type stars \citep{Osterbrock:2006}. The latter are clouds of gas formed by the ejections of matter from low mass stars between $0.8 M_\odot$ to $8 M_\odot$ in their final stage of evolution, being ionized by the hot stellar remnant \citep{Osterbrock:2006}.

H~{\sc ii} regions reflect the current chemical composition of the interstellar medium (ISM) \citep{Shaver:1983, arellano-cordova:2020, Mendez-Delgado:2022a, Groves:2023}, while PNe show the composition of the ISM at the time of the formation of the original star, along with nucleosynthetic changes that occurred during its lifetime \citep{Pequignot:2000, Leisy:2006, Karakas:2014, Delgado-Inglada:2015}. These characteristics make them very useful for complementary analysis of the distribution of their chemical composition in galaxies. \citep{Magrini:2006, Esteban:2018, Stanghellini:2018, Espinosa-Ponce:2022}.

However, there is a problem, and it is a big one: we are not certain about the correct abundance of heavy elements like O, C, N, Ne, collectively known as ``metals''. Since the pioneering works of \citet{Bowen:1939} and \citet{wyse42}, it has been known that the abundances of these elements determined from the bright collisionally excited lines (CELs) (e.g., [O~{\sc iii}] 5007, 4959) are systematically lower than those inferred from the faint recombination lines (RLs) (e.g., O~{\sc ii} 4649, 4650). The ratio between both estimates is known as the "Abundance Discrepancy Factor" (ADF), and it has been found to be around a factor of 2-4 in H~{\sc ii} regions \citep{Garcia-Rojas:2007} but can reach values of more than 500 in some PNe \citep{Wesson:2003}. The adoption of one metallicity or another can imply significant changes in the interpretation we would make about the chemical evolution of galaxies. Therefore, it is important to consider: which (if any) are the correct abundances?

\section{The favourite scenarios}

Since the emergence of this long-standing problem, several ideas have been proposed to explain the observed ADF. For instance, fluorescent excitations contaminating heavy element RLs \citep{seaton68}; errors in the atomic parameters that model the emissivity of spectral lines \citep{Rodriguez:2010}; the presence of a $\kappa$ velocity distribution in free electrons instead of a Maxwellian one \citep{Nicholls:2012}; temperature inhomogeneities \citep{peimbert67} and chemical inhomogeneities in the gas \citep{torrespeimbertetal90,Liu:2000}. 

The last two scenarios have gained particular relevance in the overall study of chemical abundances in H~{\sc ii} regions \citep{peimbert67,Peimbert:2003, ODell:2003, Esteban:2004, Garcia-Rojas:2007,stasinska:2013, Mendez-Delgado:2023a,Mendez-Delgado2023b} and PNe \citep{torrespeimbertetal90, Liu:2000,Liu:2006, Fang:2011,Garcia-Rojas:2022}. Both scenarios are supported by solid theoretical foundations, but the observational evidence to accept or discard either of these scenarios (or both) remains a matter of debate.

\subsection{Temperature inhomogeneities}
\label{sec:temps_inh}

To explain differences between the values obtained from several temperature diagnostics in ionized nebulae, \citet{peimbert67} proposed the existence of temperature inhomogeneities within the ionized gas. Since the CEL-emissivities have an exponential dependency on temperature \citep{Osterbrock:2006}, in the case of internal variations, their emissions will be substantially enhanced in the hotter volumes. On the other hand,  RLs have emissivities depending on $\sim T_{\rm e} ^{-1}$, so their emission is favored in cooler regions, although with substantially smaller bias. Mathematically, this implies that the temperature inferred from the CEL intensity ratios (such as [O~{\sc iii}] 4363/5007) will be systematically higher than the average gas temperature within the volume where the CEL-emitter ion coexists (in the case of [O~{\sc iii}] 4363/5007, the ion would be O$^{2+}$). In such case, the temperature structure of each ionization volume will be characterized by two parameters: $T_{\rm 0}$, which is the average gas temperature, and $t^2$, which quantifies how inhomogeneous the temperature is. The mathematical formalism is presented in equations \eqref{eq:topeimbert} and \eqref{eq:t2peimbert}.
\begin{equation}
    \label{eq:topeimbert}
    T_0(X^{i+})=\frac{\int T_{\rm e} n_{\rm e} n(X^{i+}) dV }{\int n_{\rm e} n(X^{i+}) dV},
\end{equation}
\begin{equation}
    \label{eq:t2peimbert}
    t^2(X^{i+})=\frac{\int [T_{\rm e}-T_0(X^{i+})]^2 n_{\rm e} n(X^{i+}) dV }{T_0(X^{i+})^2\int n_{\rm e} n(X^{i+}) dV}.
\end{equation}
In these equations, in a specific volume of the nebula, $dV$, $T_{\rm e}$, $n_{\rm e}$ are the electron temperature and density and $n(X^{i+})$ is the ion's particle density. 

Under this paradigm, when the estimated temperature from CEL intensity ratios (e.g., $T_{\rm e}$([O~{\sc iii}] 4363/5007)) is adopted to calculate chemical abundances relative to H using CEL/RL intensity ratios (e.g., [O~{\sc iii}] 5007/H$\beta$), the overestimation of temperature will result in an underestimation of ionic abundances. Conversely, if chemical abundances relative to H are calculated using RL/RL intensity ratios (e.g., O~{\sc ii} V1 4649+/H$\beta$), both lines will cancel their linear dependence on temperature, and the derived ionic abundances will be correct.  

Despite the simplicity and elegance of this proposal, its adoption as a generalized explanation for the observed ADF has not been universal, and various arguments against the existence of temperature variations have been put forward. Although classical photoionization models predict the existence of temperature inhomogeneities, these are too small ($t^2\sim0.004$) \citep{Kingdon:1995, Stasinska:2001, Ercolano:2007, Ferland:2017} to account for the observed ADF in H~{\sc ii} regions ($t^2\sim0.04$) or PNe ($t^2\sim0.06$) \citep{Peimbert:2017}. An exception might be nebula models with very high metallicity, where temperature gradients are more pronounced \citep{Stasinska:2005}. Temperature inhomogeneities in the gas tend to balance out over relatively short timescales \citep{Ferland:2016}, so their existence requires an additional physical mechanism to create and maintain them. Additionally, some PNe \citep{Liu:2000, Corradi:2015, Garcia-Rojas:2022} exhibit ADF values that are too large and appear irreconcilable with the $t^2$ formalism.

\begin{figure}[h]  
  \centering
  \includegraphics[width=0.8\linewidth]{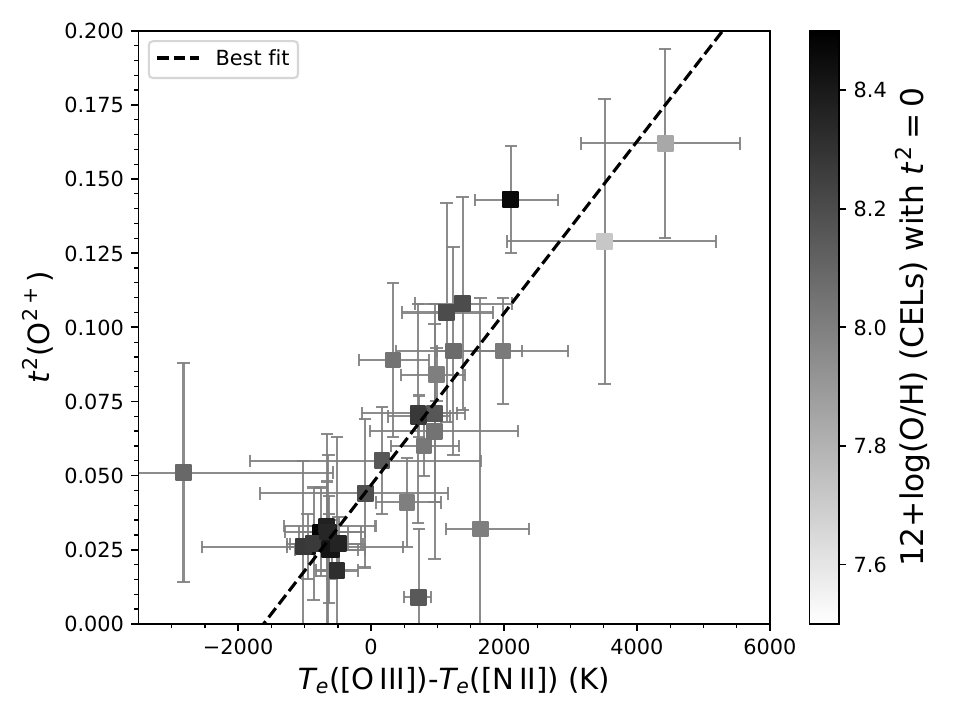}
  \caption{Relationship between $t^2(\text{O}^{2+})$, measured from O~{\sc ii} RLs and [O~{\sc iii}] CELs, with $T_{\rm e}$([O~{\sc iii}] 4363/5007) and $T_{\rm e}$([N~{\sc ii}] 5755/6584) in a sample of 32 extragalactic H~{\sc ii} regions, all the existing ones in the literature with reliable measurements of all the mentioned indicators. Reproduced from Fig.~1 and extended data Fig.~2 of \citet{Mendez-Delgado:2023a}.}
  \label{fig:Fig1}
\end{figure}

However, observational evidence seems to favour temperature variations as the cause of the ADF in H~{\sc ii} regions with precise measurements of O~{\sc ii} RLs. For instance, \citet{Garcia-Rojas:2007} found that the ADF of various ions in several Galactic H~{\sc ii} regions depends on the excitation energy of the atomic levels giving rise to CELs, which is one of the predictions of the formalism by \citet{peimbert67}. Recently, \citet{Mendez-Delgado:2023a} analyzed 32 spectra from extragalactic H~{\sc ii} regions, 20 from Galactic H~{\sc ii} regions, and 8 from Galactic ring nebulae with simultaneous detections of O~{\sc ii} RLs, $T_{\rm e}$([O~{\sc iii}] 4363/5007) and $T_{\rm e}$([N~{\sc ii}] 5755/6584). As shown in Fig\ref{fig:Fig1}, these authors demonstrated that $t^2(\text{O}^{2+})$, measured by comparing O~{\sc ii} RLs and [O~{\sc iii}] CELs \citep{Peimbert:2013}, is linearly correlated with the temperature difference $T_{\rm e}$([O~{\sc iii}] 4363/5007) $-$ $T_{\rm e}$([N~{\sc ii}] 5755/6584). This shows that $t^2(\text{O}^{2+})$ is indeed related to the gas temperature structure. Moreover, it shows that temperature variations are concentrated in regions of higher ionization degree.

The fact that temperature variations are concentrated in the zones of higher ionization, closer to the ionizing stars of the gas, points to stellar feedback as a possible cause of temperature variations. Phenomena such as shocks, winds, or even variations in the ionizing spectrum of stars can cause significant inhomogeneities in the temperature of the surrounding gas \citep{Peimbert:1991, Perez:1997, Ercolano:2007}. These are not entirely understood physical phenomena, and their incorporation into photoionization models is not widespread. \citet{Mendez-Delgado:2023a} then analyzed 8 spectra of ring nebulae, bubbles of ionized gas created by the ejection of material from young massive stars. Five of these analyzed spectra are from NGC\,6888, a nebula associated with a Wolf–Rayet WN6 star, and three others from NGC\,7635 associated with an O6.5(n)fp star \citep{esteban16}. The authors found that the observed trend between $t^2(\text{O}^{2+})$ and $T_{\rm e}$([O~{\sc iii}] 4363/5007) $-$ $T_{\rm e}$([N~{\sc ii}] 5755/6584) in the ring nebulae is very similar to what is observed in H~{\sc ii} regions. Interestingly, the spectra of NGC\,6888 exhibit the most extreme temperature variations ($t^2(\text{O}^{2+})$ $>$ 0.08), while the values for NGC\,7635 are lower ($t^2(\text{O}^{2+})$ $<$  0.06).

\begin{figure}[h]  
  \centering
  \includegraphics[width=0.8\linewidth]{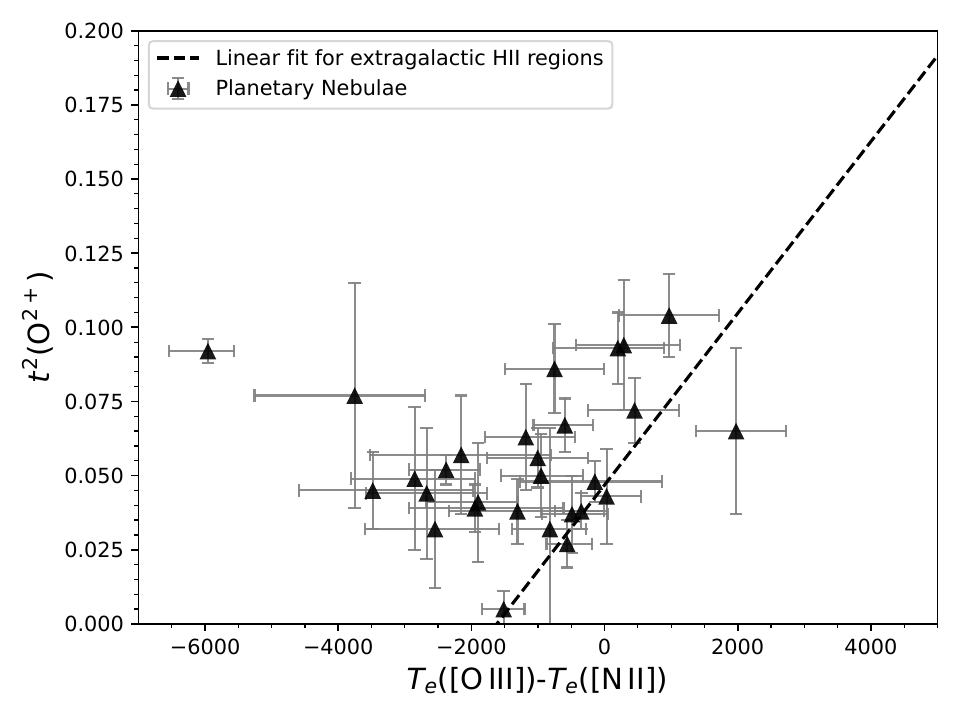}
  \caption{Figure analogous to Figure~\ref{fig:Fig1} but considering Galactic PNe. Reproduced from Fig.~10 (ArXiv version) of \citet{Mendez-Delgado:2023a}.}
  \label{fig:Fig2}
\end{figure}

However, \citet{Mendez-Delgado:2023a} tested this same hypothesis in Galactic PNe, and these do not exhibit a similar correlation as in H~{\sc ii} regions, as shown in Fig.~\ref{fig:Fig2}. This indicates that the cause of the ADF in H~{\sc ii} regions and PNe is different. Although obvious, it is important to note that even if temperature variations alone are not able to explain the observed ADF in PNe, this does not rule out their presence and impact on the determination of chemical abundances in these objects. The ADF in PNe may be caused by various linked physical phenomena. For example, the observational evidence presented by \citet{Richer:2022} for NGC 6153 shows that while most of the ADF in this region may be due to chemical inhomogeneities, temperature variations are not negligible, finding values of $t^2(\text{O}^{2+}) \approx 0.03$, which is typically found in H~{\sc ii} regions.

\subsection{Chemical inhomogeneities}
\label{sec:chems_inh}

An alternative explanation to the ADF is the existence of chemical inhomogeneities. If in a gas with a typical nebular temperature of $\sim 10000 $K there are gas components of different metallicity, they will induce temperature variations \citep{Zhang:2007}. However, if the metallicity of these inclusions is extremely high compared to their hydrogen abundance, their temperature can drop to a few thousand kelvin or even lower. Under such circumstances, the flux of O~{\sc ii} RLs may surpass the flux of the optical [O~{\sc iii}] CELs \citep{Liu:2003}, and a different mathematical formalism than that proposed by \citet{peimbert67} must be applied (note the $dV$ dependence on Eq.~\eqref{eq:t2peimbert}), as the emission from RLs and CELs will essentially come from different volumes \citep{stasinska:2002}.

The existence of these high-metallicity clumps has been particularly important to explain cases of extreme ADF \citep{Liu:2000, Yuan:2011, Corradi:2015, gomez-llanos:2020, Garcia-Rojas:2022} that have been associated with the existence of central binary stellar systems \citep{Corradi:2015, Jones:2016, Wesson:2018}. Narrow-band imaging 
 and Integral Field Unit (IFU) observations of Galactic PNe have been particularly crucial \citep{Garcia-Rojas:2016, Ali:2019, Garcia-Rojas:2022} as they have revealed morphological differences in the distribution of heavy-element RLs and their optical CEL counterparts. In addition, there are indications of kinematical differences between [O~{\sc iii}] optical CELs and O~{\sc ii} RLs \citep{Richer:2013, Richer:2022, pena:2017}, which strengthens this scenario. 

However, although this scenario provides a plausible explanation for the extreme ADF in some PNe, it does not entirely answer the initial question of which of the abundances, based on CELs or RLs, is correct. One of the most interesting proposals to assess how important the error in chemical abundance determinations is under this scenario is the "Abundance Contrast Factor (ACF)" formalism \citep[see][and the proceeding by Morisset et al. in this series]{gomez-llanos:2020}. The ACF measures the proportion of flux emitted by the cold component relative to the total emission. If the cold component is extremely hydrogen-poor, heavy-element RLs will have the most significant errors. However, H~{\sc i} RLs can also have a significant bias by increasing their emission in the cold component, affecting the inferred abundances from heavy-element CEL to H~{\sc i} RL intensity ratios.

Even in cases where the high-metallicity clumps are so hydrogen-poor that the impact on H~{\sc i} RLs is negligible, abundances observationally determined with CELs may not be reliable. Besides potential existence of real temperature inhomogeneities in the ``hot'' gas component emitting CELs, the existence of extremely cold clumps emitting a large amount of heavy-element RLs could bias the $T_{\rm e}$ values inferred from auroral to nebular CEL ratios by contaminating the emission of auroral lines (e.g., [O~{\sc iii}] $\lambda$4363, [N~{\sc ii}] $\lambda$5755) \citep{Rubin:1986, gomez-llanos2020b}. In order to ``decontaminate" the temperatures determined with CEL ratios, it is necessary to know precisely the temperature and ionic abundances of the high-metallicity clump that is emitting the ``RL-contamination''. This requires determining the physical conditions using heavy-element RLs such as O~{\sc ii} $\lambda$4089/$\lambda$4649 to determine $T_{\rm e}$ or O~{\sc ii} $\lambda$4661/$\lambda$4649 to determine $n_{\rm e}$ \citep{Wesson:2005, Peimbert:2013, Storey:2017}. This involves measuring heavy-element RLs that are comparatively fainter than those traditionally used to determine chemical abundances. A correction like the one proposed by \citet{Liu:2000} in their equations (1) and (2) using physical conditions and ionic abundances determined in the ``hot'' gas component is conceptually incorrect, and it has not been demonstrated to be ``less incorrect'' than not making any correction.

\begin{figure}[h]  
  \centering
  \includegraphics[width=0.8\linewidth]{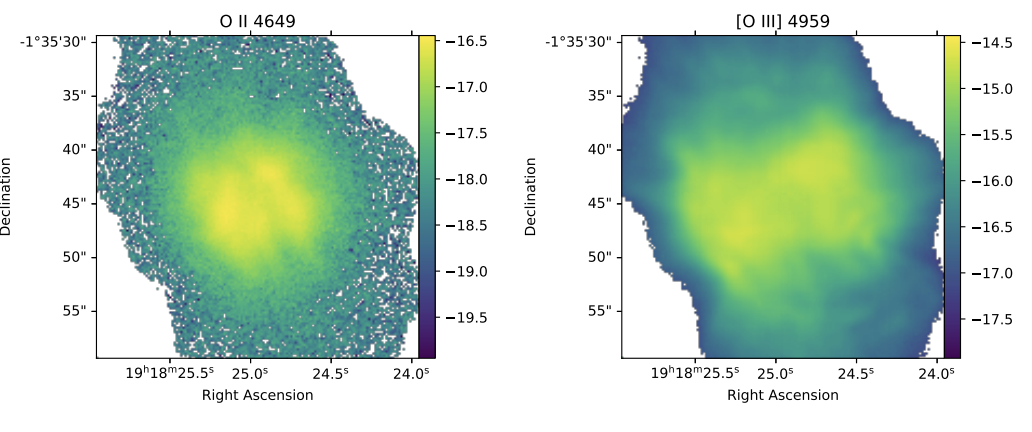}
  \caption{ MUSE distribution of the emission of O~{\sc ii} 4649 and [O~{\sc iii}] $\lambda$4959 in the high-ADF Galactic planetary nebula NGC\,6778 taken from \citet{Garcia-Rojas:2022}. The morphological differences are evident and have been interpreted as evidence of the existence of components with different metallicity in the nebula.}
  \label{fig:Fig3}
\end{figure}

In addition to the difficulty of explaining the creation and survival of high-metallicity clumps \citep{Henney:2010}, extrapolating the existence of chemical inhomogeneities to explain the ADF in all PNe, not just extreme cases, could imply the existence of a wide variety of ACF values. If the high-metallicity component is not so hydrogen-poor, the emission of both CELs and RLs can come from both volumes. Such a scenario would imply that both estimates of CELs and RLs using the standard ``direct method'' are essentially incorrect. On the other hand, there are still some interesting questions about the observational evidence of the high-metallicity components. For example, \citet{Garcia-Rojas:2022} has shown that the emission of O~{\sc ii} RLs has a different morphology than its collisionally excited counterparts, as shown in Fig. \ref{fig:Fig3}. This has been argued as evidence of the presence of a cold gas component with a temperature of $\sim 800$ K. However, excellent MUSE maps show that highly ionized CELs such as [Cl~{\sc iv}] or [Ar~IV] have a similar distribution to O~{\sc ii} RLs. This does not seem to be a contrast problem since weaker lines than O~{\sc ii} $\lambda$4649, such as [Cl~{\sc iii}] $\lambda$5518, have a distribution similar to that of [O~{\sc iii}] CELs. Since both Cl and Ar are third-row elements in the periodic table, this also does not seem to be a recombination contribution to these CELs \citep{Barlow:2003}. How could a clump with a temperature of $\sim 800$ K could emit these highly ionized CELs?

\begin{figure}[h]  
  \centering
  \includegraphics[width=0.4\linewidth]{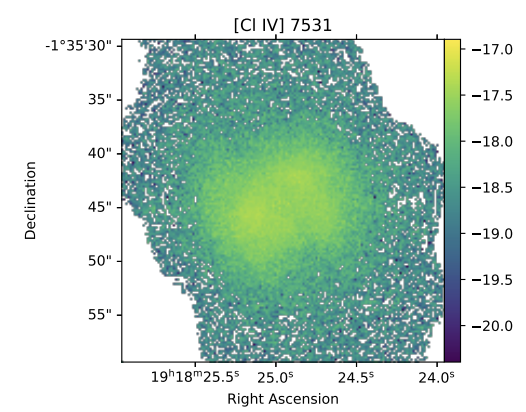}
  \includegraphics[width=0.4\linewidth]{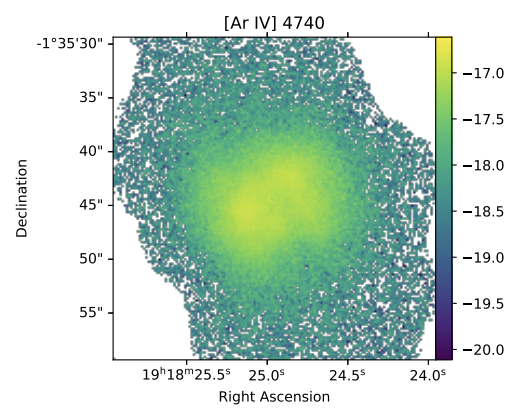}
  \caption{ MUSE distribution of the emission of [Cl~{\sc iv}] $\lambda$7531 and [Ar~{\sc iv}] $\lambda$4740 in the high-ADF Galactic planetary nebula NGC\,6778 taken from \citet{Garcia-Rojas:2022}. The morphology of these highly ionized CELs are more similar to those of heavy element RLs (see Fig.\ref{fig:Fig3}).}
  \label{fig:Fig4}
\end{figure}

\section{Conclusions}

The systematic difference between heavy element abundances determined with RLs and CELs, the so-called abundance discrepancy problem, is one of the most interesting and important enigmas in the study of the chemical composition of ionized nebulae. This long-standing problem has significant implications for models of galaxy formation and evolution, and until its complete clarification, these models may be called into question. However, over the years, significant progress has been made thanks to deep observations of ionized nebulae collected over several decades by numerous researchers worldwide.

The evidence suggests that the ADF may have different origins in H~{\sc ii} regions and PNe. Recent observational studies seem to favor the presence of temperature inhomogeneities in highly ionized gas as the cause of the ADF in H~{\sc ii} regions, in consistency with the formalism of \citet{peimbert67}. In this case, heavy element RLs appear to provide the correct chemical abundances, being 2$-$4 times higher than those obtained using CELs with the ``direct method'' \citep[which assumes homogeneous electron temperature, see][]{Mendez-Delgado:2023a}.

The case of PNe appears more intriguing. The evidence suggests that the presence of temperature variations alone is not able to quantitatively explain the observed ADFs in many PNe. In these regions, especially in those with the highest ADF values, there are indications that components of gas with different metallicity could exist. In fact, narrow-band imaging and IFU observations reveal that heavy-element RLs have morphological distributions different from their collisionally excited counterparts, which is qualitatively consistent with this scenario. However, the quantitative impact of the presence of high-metallicity clumps in the gas remains an open question which is now being addressed by several research groups \citep{gomez-llanos:2020, Garcia-Rojas:2022}. Finally, although the differentiated morphological distribution between heavy-element RLs and CELs is evident in PNe with high ADFs, it is important to understand exactly why this is also observed in highly ionized CELs such as [Cl~{\sc iv}] or [Ar~{\sc iv}]. More detailed IFU observations of PNe considering the morphological distribution of RLs and CELs of more ions will shed light on this interesting problem.\\

JEMD thanks the organizers for their invitation to this excellent symposium as an invited speaker. JEMD gratefully acknowledge funding from the Deutsche Forschungsgemeinschaft (DFG, German Research Foundation) in the form of an Emmy Noether Research Group (grant number - KR4598/2-1, PI Kreckel). JG-R acknowledges financial support from the Canarian Agency for Research, Innovation and Information Society (ACIISI), of the Canary Islands Government, and the European Regional Development Fund (ERDF), under grant with reference ProID2021010074, and from grant P/308614 financed by funds transferred from the Spanish Ministry of Science, Innovation and Universities, charged to the General State Budgets and with funds transferred from the General Budgets of the Autonomous Community of the Canary Islands by the MCIU. JG-R also acknowledges funds from the Spanish Ministry of Science and Innovation (MICINN) through the Spanish State Research Agency, under Severo Ochoa Centres of Excellence Programme 2020-2023 (CEX2019-000920-S).

\bibliographystyle{mnras}
\bibliography{iauguide}

\begin{thebibliography}{}
\makeatletter
\relax
\def\mn@urlcharsother{\let\do\@makeother \do\$\do\&\do\#\do\^\do\_\do\%\do\~}
\def\mn@doi{\begingroup\mn@urlcharsother \@ifnextchar [ {\mn@doi@}
  {\mn@doi@[]}}
\def\mn@doi@[#1]#2{\def\@tempa{#1}\ifx\@tempa\@empty \href
  {http://dx.doi.org/#2} {doi:#2}\else \href {http://dx.doi.org/#2} {#1}\fi
  \endgroup}
\def\mn@eprint#1#2{\mn@eprint@#1:#2::\@nil}
\def\mn@eprint@arXiv#1{\href {http://arxiv.org/abs/#1} {{\tt arXiv:#1}}}
\def\mn@eprint@dblp#1{\href {http://dblp.uni-trier.de/rec/bibtex/#1.xml}
  {dblp:#1}}
\def\mn@eprint@#1:#2:#3:#4\@nil{\def\@tempa {#1}\def\@tempb {#2}\def\@tempc
  {#3}\ifx \@tempc \@empty \let \@tempc \@tempb \let \@tempb \@tempa \fi \ifx
  \@tempb \@empty \def\@tempb {arXiv}\fi \@ifundefined
  {mn@eprint@\@tempb}{\@tempb:\@tempc}{\expandafter \expandafter \csname
  mn@eprint@\@tempb\endcsname \expandafter{\@tempc}}}

\bibitem[\protect\citeauthoryear{{Ali} \& {Dopita}}{{Ali} \&
  {Dopita}}{2019}]{Ali:2019}
{Ali} A.,  {Dopita} M.~A.,  2019, \mn@doi [\mnras] {10.1093/mnras/stz201},
  \href {https://ui.adsabs.harvard.edu/abs/2019MNRAS.484.3251A} {484, 3251}

\bibitem[\protect\citeauthoryear{{Arellano-C{\'o}rdova}, {Esteban},
  {Garc{\'\i}a-Rojas}  \& {M{\'e}ndez-Delgado}}{{Arellano-C{\'o}rdova}
  et~al.}{2020}]{arellano-cordova:2020}
{Arellano-C{\'o}rdova} K.~Z.,  {Esteban} C.,  {Garc{\'\i}a-Rojas} J.,
  {M{\'e}ndez-Delgado} J.~E.,  2020, \mn@doi [\mnras] {10.1093/mnras/staa1523},
  \href {https://ui.adsabs.harvard.edu/abs/2020MNRAS.496.1051A} {496, 1051}

\bibitem[\protect\citeauthoryear{{Barlow}, {Liu}, {. P{\'e}quignot}, {Storey},
  {Tsamis}  \& {Morisset}}{{Barlow} et~al.}{2003}]{Barlow:2003}
{Barlow} M.~J.,  {Liu} X.~W.,  {. P{\'e}quignot} D.,  {Storey} P.~J.,  {Tsamis}
  Y.~G.,   {Morisset} C.,  2003, in {Kwok} S.,  {Dopita} M.,   {Sutherland} R.,
   eds, ~ Vol. 209, Planetary Nebulae: Their Evolution and Role in the
  Universe. p.~373

\bibitem[\protect\citeauthoryear{{Bethe}}{{Bethe}}{1939}]{Bethe:1939}
{Bethe} H.~A.,  1939, \mn@doi [Physical Review] {10.1103/PhysRev.55.434}, \href
  {https://ui.adsabs.harvard.edu/abs/1939PhRv...55..434B} {55, 434}

\bibitem[\protect\citeauthoryear{{Bowen} \& {Wyse}}{{Bowen} \&
  {Wyse}}{1939}]{Bowen:1939}
{Bowen} I.~S.,  {Wyse} A.~B.,  1939, \mn@doi [Lick Observatory Bulletin]
  {10.5479/ADS/bib/1939LicOB.19.1B}, \href
  {https://ui.adsabs.harvard.edu/abs/1939LicOB..19....1B} {495, 1}

\bibitem[\protect\citeauthoryear{{Caughlan}}{{Caughlan}}{1965}]{Caughlan:1965}
{Caughlan} G.~R.,  1965, \mn@doi [\apj] {10.1086/148155}, \href
  {https://ui.adsabs.harvard.edu/abs/1965ApJ...141..688C} {141, 688}

\bibitem[\protect\citeauthoryear{{Corradi}, {Garc{\'\i}a-Rojas}, {Jones}  \&
  {Rodr{\'\i}guez-Gil}}{{Corradi} et~al.}{2015}]{Corradi:2015}
{Corradi} R. L.~M.,  {Garc{\'\i}a-Rojas} J.,  {Jones} D.,
  {Rodr{\'\i}guez-Gil} P.,  2015, \mn@doi [\apj] {10.1088/0004-637X/803/2/99},
  \href {https://ui.adsabs.harvard.edu/abs/2015ApJ...803...99C} {803, 99}

\bibitem[\protect\citeauthoryear{{Delgado-Inglada}, {Rodr{\'\i}guez},
  {Peimbert}, {Stasi{\'n}ska}  \& {Morisset}}{{Delgado-Inglada}
  et~al.}{2015}]{Delgado-Inglada:2015}
{Delgado-Inglada} G.,  {Rodr{\'\i}guez} M.,  {Peimbert} M.,  {Stasi{\'n}ska}
  G.,   {Morisset} C.,  2015, \mn@doi [\mnras] {10.1093/mnras/stv388}, \href
  {https://ui.adsabs.harvard.edu/abs/2015MNRAS.449.1797D} {449, 1797}

\bibitem[\protect\citeauthoryear{{Ercolano}, {Bastian}  \&
  {Stasi{\'n}ska}}{{Ercolano} et~al.}{2007}]{Ercolano:2007}
{Ercolano} B.,  {Bastian} N.,   {Stasi{\'n}ska} G.,  2007, \mn@doi [\mnras]
  {10.1111/j.1365-2966.2007.12002.x}, \href
  {https://ui.adsabs.harvard.edu/abs/2007MNRAS.379..945E} {379, 945}

\bibitem[\protect\citeauthoryear{{Espinosa-Ponce}, {S{\'a}nchez}, {Morisset},
  {Barrera-Ballesteros}, {Galbany}, {Garc{\'\i}a-Benito}, {Lacerda}  \&
  {Mast}}{{Espinosa-Ponce} et~al.}{2022}]{Espinosa-Ponce:2022}
{Espinosa-Ponce} C.,  {S{\'a}nchez} S.~F.,  {Morisset} C.,
  {Barrera-Ballesteros} J.~K.,  {Galbany} L.,  {Garc{\'\i}a-Benito} R.,
  {Lacerda} E.~A.~D.,   {Mast} D.,  2022, \mn@doi [\mnras]
  {10.1093/mnras/stac456}, \href
  {https://ui.adsabs.harvard.edu/abs/2022MNRAS.512.3436E} {512, 3436}

\bibitem[\protect\citeauthoryear{{Esteban} \& {Garc{\'\i}a-Rojas}}{{Esteban} \&
  {Garc{\'\i}a-Rojas}}{2018}]{Esteban:2018}
{Esteban} C.,  {Garc{\'\i}a-Rojas} J.,  2018, \mn@doi [\mnras]
  {10.1093/mnras/sty1168}, \href
  {https://ui.adsabs.harvard.edu/abs/2018MNRAS.478.2315E} {478, 2315}

\bibitem[\protect\citeauthoryear{{Esteban}, {Peimbert}, {Garc{\'\i}a-Rojas},
  {Ruiz}, {Peimbert}  \& {Rodr{\'\i}guez}}{{Esteban}
  et~al.}{2004}]{Esteban:2004}
{Esteban} C.,  {Peimbert} M.,  {Garc{\'\i}a-Rojas} J.,  {Ruiz} M.~T.,
  {Peimbert} A.,   {Rodr{\'\i}guez} M.,  2004, \mn@doi [\mnras]
  {10.1111/j.1365-2966.2004.08313.x}, \href
  {https://ui.adsabs.harvard.edu/abs/2004MNRAS.355..229E} {355, 229}

\bibitem[\protect\citeauthoryear{{Esteban}, {Mesa-Delgado}, {Morisset}  \&
  {Garc{\'\i}a-Rojas}}{{Esteban} et~al.}{2016}]{esteban16}
{Esteban} C.,  {Mesa-Delgado} A.,  {Morisset} C.,   {Garc{\'\i}a-Rojas} J.,
  2016, \mn@doi [\mnras] {10.1093/mnras/stw1243}, \href
  {https://ui.adsabs.harvard.edu/abs/2016MNRAS.460.4038E} {460, 4038}

\bibitem[\protect\citeauthoryear{{Fang} \& {Liu}}{{Fang} \&
  {Liu}}{2011}]{Fang:2011}
{Fang} X.,  {Liu} X.~W.,  2011, \mn@doi [\mnras]
  {10.1111/j.1365-2966.2011.18681.x}, \href
  {https://ui.adsabs.harvard.edu/abs/2011MNRAS.415..181F} {415, 181}

\bibitem[\protect\citeauthoryear{{Ferland}, {Henney}, {O'Dell}  \&
  {Peimbert}}{{Ferland} et~al.}{2016}]{Ferland:2016}
{Ferland} G.~J.,  {Henney} W.~J.,  {O'Dell} C.~R.,   {Peimbert} M.,  2016,
  \mn@doi [\rmxaa] {10.48550/arXiv.1605.03634}, \href
  {https://ui.adsabs.harvard.edu/abs/2016RMxAA..52..261F} {52, 261}

\bibitem[\protect\citeauthoryear{{Ferland} et~al.,}{{Ferland}
  et~al.}{2017}]{Ferland:2017}
{Ferland} G.~J.,  et~al., 2017, \mn@doi [\rmxaa] {10.48550/arXiv.1705.10877},
  \href {https://ui.adsabs.harvard.edu/abs/2017RMxAA..53..385F} {53, 385}

\bibitem[\protect\citeauthoryear{{Garc{\'\i}a-Rojas} \&
  {Esteban}}{{Garc{\'\i}a-Rojas} \& {Esteban}}{2007}]{Garcia-Rojas:2007}
{Garc{\'\i}a-Rojas} J.,  {Esteban} C.,  2007, \mn@doi [\apj] {10.1086/521871},
  \href {https://ui.adsabs.harvard.edu/abs/2007ApJ...670..457G} {670, 457}

\bibitem[\protect\citeauthoryear{{Garc{\'\i}a-Rojas}, {Corradi}, {Monteiro},
  {Jones}, {Rodr{\'\i}guez-Gil}  \& {Cabrera-Lavers}}{{Garc{\'\i}a-Rojas}
  et~al.}{2016}]{Garcia-Rojas:2016}
{Garc{\'\i}a-Rojas} J.,  {Corradi} R. L.~M.,  {Monteiro} H.,  {Jones} D.,
  {Rodr{\'\i}guez-Gil} P.,   {Cabrera-Lavers} A.,  2016, \mn@doi [\apjl]
  {10.3847/2041-8205/824/2/L27}, \href
  {https://ui.adsabs.harvard.edu/abs/2016ApJ...824L..27G} {824, L27}

\bibitem[\protect\citeauthoryear{{Garc{\'\i}a-Rojas}, {Morisset}, {Jones},
  {Wesson}, {Boffin}, {Monteiro}, {Corradi}  \&
  {Rodr{\'\i}guez-Gil}}{{Garc{\'\i}a-Rojas} et~al.}{2022}]{Garcia-Rojas:2022}
{Garc{\'\i}a-Rojas} J.,  {Morisset} C.,  {Jones} D.,  {Wesson} R.,  {Boffin}
  H.~M.~J.,  {Monteiro} H.,  {Corradi} R.~L.~M.,   {Rodr{\'\i}guez-Gil} P.,
  2022, \mn@doi [\mnras] {10.1093/mnras/stab3523}, \href
  {https://ui.adsabs.harvard.edu/abs/2022MNRAS.510.5444G} {510, 5444}

\bibitem[\protect\citeauthoryear{{G{\'o}mez-Llanos} \&
  {Morisset}}{{G{\'o}mez-Llanos} \& {Morisset}}{2020}]{gomez-llanos:2020}
{G{\'o}mez-Llanos} V.,  {Morisset} C.,  2020, \mn@doi [\mnras]
  {10.1093/mnras/staa2157}, \href
  {https://ui.adsabs.harvard.edu/abs/2020MNRAS.497.3363G} {497, 3363}

\bibitem[\protect\citeauthoryear{{G{\'o}mez-Llanos}, {Morisset},
  {Garc{\'\i}a-Rojas}, {Jones}, {Wesson}, {Corradi}  \&
  {Boffin}}{{G{\'o}mez-Llanos} et~al.}{2020}]{gomez-llanos2020b}
{G{\'o}mez-Llanos} V.,  {Morisset} C.,  {Garc{\'\i}a-Rojas} J.,  {Jones} D.,
  {Wesson} R.,  {Corradi} R.~L.~M.,   {Boffin} H.~M.~J.,  2020, \mn@doi
  [\mnras] {10.1093/mnrasl/slaa131}, \href
  {https://ui.adsabs.harvard.edu/abs/2020MNRAS.498L..82G} {498, L82}

\bibitem[\protect\citeauthoryear{{Groves} et~al.,}{{Groves}
  et~al.}{2023}]{Groves:2023}
{Groves} B.,  et~al., 2023, \mn@doi [\mnras] {10.1093/mnras/stad114}, \href
  {https://ui.adsabs.harvard.edu/abs/2023MNRAS.520.4902G} {520, 4902}

\bibitem[\protect\citeauthoryear{{Henney} \& {Stasi{\'n}ska}}{{Henney} \&
  {Stasi{\'n}ska}}{2010}]{Henney:2010}
{Henney} W.~J.,  {Stasi{\'n}ska} G.,  2010, \mn@doi [\apj]
  {10.1088/0004-637X/711/2/881}, \href
  {https://ui.adsabs.harvard.edu/abs/2010ApJ...711..881H} {711, 881}

\bibitem[\protect\citeauthoryear{{Jones}, {Wesson}, {Garc{\'\i}a-Rojas},
  {Corradi}  \& {Boffin}}{{Jones} et~al.}{2016}]{Jones:2016}
{Jones} D.,  {Wesson} R.,  {Garc{\'\i}a-Rojas} J.,  {Corradi} R.~L.~M.,
  {Boffin} H.~M.~J.,  2016, \mn@doi [\mnras] {10.1093/mnras/stv2519}, \href
  {https://ui.adsabs.harvard.edu/abs/2016MNRAS.455.3263J} {455, 3263}

\bibitem[\protect\citeauthoryear{{Karakas} \& {Lattanzio}}{{Karakas} \&
  {Lattanzio}}{2014}]{Karakas:2014}
{Karakas} A.~I.,  {Lattanzio} J.~C.,  2014, \mn@doi [\pasa]
  {10.1017/pasa.2014.21}, \href
  {https://ui.adsabs.harvard.edu/abs/2014PASA...31...30K} {31, e030}

\bibitem[\protect\citeauthoryear{{Kingdon} \& {Ferland}}{{Kingdon} \&
  {Ferland}}{1995}]{Kingdon:1995}
{Kingdon} J.~B.,  {Ferland} G.~J.,  1995, \mn@doi [\apj] {10.1086/176175},
  \href {https://ui.adsabs.harvard.edu/abs/1995ApJ...450..691K} {450, 691}

\bibitem[\protect\citeauthoryear{{Leisy} \& {Dennefeld}}{{Leisy} \&
  {Dennefeld}}{2006}]{Leisy:2006}
{Leisy} P.,  {Dennefeld} M.,  2006, \mn@doi [\aap]
  {10.1051/0004-6361:20053063}, \href
  {https://ui.adsabs.harvard.edu/abs/2006A&A...456..451L} {456, 451}

\bibitem[\protect\citeauthoryear{{Liu}}{{Liu}}{2003}]{Liu:2003}
{Liu} X.~W.,  2003, in {Kwok} S.,  {Dopita} M.,   {Sutherland} R.,  eds, ~ Vol.
  209, Planetary Nebulae: Their Evolution and Role in the Universe. p.~339

\bibitem[\protect\citeauthoryear{{Liu}, {Storey}, {Barlow}, {Danziger}, {Cohen}
   \& {Bryce}}{{Liu} et~al.}{2000}]{Liu:2000}
{Liu} X.~W.,  {Storey} P.~J.,  {Barlow} M.~J.,  {Danziger} I.~J.,  {Cohen} M.,
   {Bryce} M.,  2000, \mn@doi [\mnras] {10.1046/j.1365-8711.2000.03167.x},
  \href {https://ui.adsabs.harvard.edu/abs/2000MNRAS.312..585L} {312, 585}

\bibitem[\protect\citeauthoryear{{Liu}, {Barlow}, {Zhang}, {Bastin}  \&
  {Storey}}{{Liu} et~al.}{2006}]{Liu:2006}
{Liu} X.~W.,  {Barlow} M.~J.,  {Zhang} Y.,  {Bastin} R.~J.,   {Storey} P.~J.,
  2006, \mn@doi [\mnras] {10.1111/j.1365-2966.2006.10283.x}, \href
  {https://ui.adsabs.harvard.edu/abs/2006MNRAS.368.1959L} {368, 1959}

\bibitem[\protect\citeauthoryear{{Magrini}, {Coccato}, {Stanghellini},
  {Casasola}  \& {Galli}}{{Magrini} et~al.}{2016}]{Magrini:2006}
{Magrini} L.,  {Coccato} L.,  {Stanghellini} L.,  {Casasola} V.,   {Galli} D.,
  2016, \mn@doi [\aap] {10.1051/0004-6361/201527799}, \href
  {https://ui.adsabs.harvard.edu/abs/2016A&A...588A..91M} {588, A91}

\bibitem[\protect\citeauthoryear{{Maiolino} \& {Mannucci}}{{Maiolino} \&
  {Mannucci}}{2019}]{Maiolino:2019}
{Maiolino} R.,  {Mannucci} F.,  2019, \mn@doi [\aapr]
  {10.1007/s00159-018-0112-2}, \href
  {https://ui.adsabs.harvard.edu/abs/2019A&ARv..27....3M} {27, 3}

\bibitem[\protect\citeauthoryear{{M{\'e}ndez-Delgado}, {Amayo},
  {Arellano-C{\'o}rdova}, {Esteban}, {Garc{\'\i}a-Rojas}, {Carigi}  \&
  {Delgado-Inglada}}{{M{\'e}ndez-Delgado} et~al.}{2022}]{Mendez-Delgado:2022a}
{M{\'e}ndez-Delgado} J.~E.,  {Amayo} A.,  {Arellano-C{\'o}rdova} K.~Z.,
  {Esteban} C.,  {Garc{\'\i}a-Rojas} J.,  {Carigi} L.,   {Delgado-Inglada} G.,
  2022, \mn@doi [\mnras] {10.1093/mnras/stab3782}, \href
  {https://ui.adsabs.harvard.edu/abs/2022MNRAS.510.4436M} {510, 4436}

\bibitem[\protect\citeauthoryear{{M{\'e}ndez-Delgado}
  et~al.,}{{M{\'e}ndez-Delgado} et~al.}{2023a}]{Mendez-Delgado2023b}
{M{\'e}ndez-Delgado} J.~E.,  et~al., 2023a, \mn@doi [\mnras]
  {10.1093/mnras/stad1569}, \href
  {https://ui.adsabs.harvard.edu/abs/2023MNRAS.523.2952M} {523, 2952}

\bibitem[\protect\citeauthoryear{{M{\'e}ndez-Delgado}, {Esteban},
  {Garc{\'\i}a-Rojas}, {Kreckel}  \& {Peimbert}}{{M{\'e}ndez-Delgado}
  et~al.}{2023b}]{Mendez-Delgado:2023a}
{M{\'e}ndez-Delgado} J.~E.,  {Esteban} C.,  {Garc{\'\i}a-Rojas} J.,  {Kreckel}
  K.,   {Peimbert} M.,  2023b, \mn@doi [\nat] {10.1038/s41586-023-05956-2},
  \href {https://ui.adsabs.harvard.edu/abs/2023Natur.618..249M} {618, 249}

\bibitem[\protect\citeauthoryear{{Nicholls}, {Dopita}  \&
  {Sutherland}}{{Nicholls} et~al.}{2012}]{Nicholls:2012}
{Nicholls} D.~C.,  {Dopita} M.~A.,   {Sutherland} R.~S.,  2012, \mn@doi [\apj]
  {10.1088/0004-637X/752/2/148}, \href
  {https://ui.adsabs.harvard.edu/abs/2012ApJ...752..148N} {752, 148}

\bibitem[\protect\citeauthoryear{{O'Dell}, {Peimbert}  \& {Peimbert}}{{O'Dell}
  et~al.}{2003}]{ODell:2003}
{O'Dell} C.~R.,  {Peimbert} M.,   {Peimbert} A.,  2003, \mn@doi [\aj]
  {10.1086/374788}, \href
  {https://ui.adsabs.harvard.edu/abs/2003AJ....125.2590O} {125, 2590}

\bibitem[\protect\citeauthoryear{{Osterbrock} \& {Ferland}}{{Osterbrock} \&
  {Ferland}}{2006}]{Osterbrock:2006}
{Osterbrock} D.~E.,  {Ferland} G.~J.,  2006, {Astrophysics of gaseous nebulae
  and active galactic nuclei}

\bibitem[\protect\citeauthoryear{{Pe{\~n}a}, {Ruiz-Escobedo},
  {Rechy-Garc{\'\i}a}  \& {Garc{\'\i}a-Rojas}}{{Pe{\~n}a}
  et~al.}{2017}]{pena:2017}
{Pe{\~n}a} M.,  {Ruiz-Escobedo} F.,  {Rechy-Garc{\'\i}a} J.~S.,
  {Garc{\'\i}a-Rojas} J.,  2017, \mn@doi [\mnras] {10.1093/mnras/stx1991},
  \href {https://ui.adsabs.harvard.edu/abs/2017MNRAS.472.1182P} {472, 1182}

\bibitem[\protect\citeauthoryear{{Peebles}}{{Peebles}}{1966a}]{Peebles:1966}
{Peebles} P.~J.,  1966a, \mn@doi [\prl] {10.1103/PhysRevLett.16.410}, \href
  {https://ui.adsabs.harvard.edu/abs/1966PhRvL..16..410P} {16, 410}

\bibitem[\protect\citeauthoryear{{Peebles}}{{Peebles}}{1966b}]{Peebles:1966b}
{Peebles} P.~J.~E.,  1966b, \mn@doi [\apj] {10.1086/148918}, \href
  {https://ui.adsabs.harvard.edu/abs/1966ApJ...146..542P} {146, 542}

\bibitem[\protect\citeauthoryear{{Peimbert}}{{Peimbert}}{1967}]{peimbert67}
{Peimbert} M.,  1967, \mn@doi [ApJ] {10.1086/149385}, \href
  {https://ui.adsabs.harvard.edu/abs/1967ApJ...150..825P} {150, 825}

\bibitem[\protect\citeauthoryear{{Peimbert}}{{Peimbert}}{2003}]{Peimbert:2003}
{Peimbert} A.,  2003, \mn@doi [\apj] {10.1086/345793}, \href
  {https://ui.adsabs.harvard.edu/abs/2003ApJ...584..735P} {584, 735}

\bibitem[\protect\citeauthoryear{{Peimbert} \& {Peimbert}}{{Peimbert} \&
  {Peimbert}}{2013}]{Peimbert:2013}
{Peimbert} A.,  {Peimbert} M.,  2013, \mn@doi [\apj]
  {10.1088/0004-637X/778/2/89}, \href
  {https://ui.adsabs.harvard.edu/abs/2013ApJ...778...89P} {778, 89}

\bibitem[\protect\citeauthoryear{{Peimbert}, {Sarmiento}  \&
  {Fierro}}{{Peimbert} et~al.}{1991}]{Peimbert:1991}
{Peimbert} M.,  {Sarmiento} A.,   {Fierro} J.,  1991, \mn@doi [\pasp]
  {10.1086/132886}, \href
  {https://ui.adsabs.harvard.edu/abs/1991PASP..103..815P} {103, 815}

\bibitem[\protect\citeauthoryear{{Peimbert}, {Peimbert}  \&
  {Delgado-Inglada}}{{Peimbert} et~al.}{2017}]{Peimbert:2017}
{Peimbert} M.,  {Peimbert} A.,   {Delgado-Inglada} G.,  2017, \mn@doi [\pasp]
  {10.1088/1538-3873/aa72c3}, \href
  {https://ui.adsabs.harvard.edu/abs/2017PASP..129h2001P} {129, 082001}

\bibitem[\protect\citeauthoryear{{P{\'e}quignot}, {Walsh}, {Zijlstra}  \&
  {Dudziak}}{{P{\'e}quignot} et~al.}{2000}]{Pequignot:2000}
{P{\'e}quignot} D.,  {Walsh} J.~R.,  {Zijlstra} A.~A.,   {Dudziak} G.,  2000,
  \mn@doi [\aap] {10.48550/arXiv.astro-ph/0008443}, \href
  {https://ui.adsabs.harvard.edu/abs/2000A&A...361L...1P} {361, L1}

\bibitem[\protect\citeauthoryear{{Perez}}{{Perez}}{1997}]{Perez:1997}
{Perez} E.,  1997, \mn@doi [\mnras] {10.1093/mnras/290.3.465}, \href
  {https://ui.adsabs.harvard.edu/abs/1997MNRAS.290..465P} {290, 465}

\bibitem[\protect\citeauthoryear{{Richer}, {Georgiev}, {Arrieta}  \&
  {Torres-Peimbert}}{{Richer} et~al.}{2013}]{Richer:2013}
{Richer} M.~G.,  {Georgiev} L.,  {Arrieta} A.,   {Torres-Peimbert} S.,  2013,
  \mn@doi [\apj] {10.1088/0004-637X/773/2/133}, \href
  {https://ui.adsabs.harvard.edu/abs/2013ApJ...773..133R} {773, 133}

\bibitem[\protect\citeauthoryear{{Richer}, {Arrieta}, {Arias},
  {Casta{\~n}eda-Carlos}, {Torres-Peimbert}, {L{\'o}pez}  \&
  {Galindo}}{{Richer} et~al.}{2022}]{Richer:2022}
{Richer} M.~G.,  {Arrieta} A.,  {Arias} L.,  {Casta{\~n}eda-Carlos} L.,
  {Torres-Peimbert} S.,  {L{\'o}pez} J.~A.,   {Galindo} A.,  2022, \mn@doi
  [\aj] {10.3847/1538-3881/ac9732}, \href
  {https://ui.adsabs.harvard.edu/abs/2022AJ....164..243R} {164, 243}

\bibitem[\protect\citeauthoryear{{Rodr{\'\i}guez} \&
  {Garc{\'\i}a-Rojas}}{{Rodr{\'\i}guez} \&
  {Garc{\'\i}a-Rojas}}{2010}]{Rodriguez:2010}
{Rodr{\'\i}guez} M.,  {Garc{\'\i}a-Rojas} J.,  2010, \mn@doi [\apj]
  {10.1088/0004-637X/708/2/1551}, \href
  {https://ui.adsabs.harvard.edu/abs/2010ApJ...708.1551R} {708, 1551}

\bibitem[\protect\citeauthoryear{{Rubin}}{{Rubin}}{1986}]{Rubin:1986}
{Rubin} R.~H.,  1986, \mn@doi [\apj] {10.1086/164606}, \href
  {https://ui.adsabs.harvard.edu/abs/1986ApJ...309..334R} {309, 334}

\bibitem[\protect\citeauthoryear{{Seaton}}{{Seaton}}{1968}]{seaton68}
{Seaton} M.~J.,  1968, \mn@doi [\mnras] {10.1093/mnras/139.2.129}, \href
  {https://ui.adsabs.harvard.edu/abs/1968MNRAS.139..129S} {139, 129}

\bibitem[\protect\citeauthoryear{{Shaver}, {McGee}, {Newton}, {Danks}  \&
  {Pottasch}}{{Shaver} et~al.}{1983}]{Shaver:1983}
{Shaver} P.~A.,  {McGee} R.~X.,  {Newton} L.~M.,  {Danks} A.~C.,   {Pottasch}
  S.~R.,  1983, \mn@doi [\mnras] {10.1093/mnras/204.1.53}, \href
  {https://ui.adsabs.harvard.edu/abs/1983MNRAS.204...53S} {204, 53}

\bibitem[\protect\citeauthoryear{{Stanghellini} \& {Haywood}}{{Stanghellini} \&
  {Haywood}}{2018}]{Stanghellini:2018}
{Stanghellini} L.,  {Haywood} M.,  2018, \mn@doi [\apj]
  {10.3847/1538-4357/aacaf8}, \href
  {https://ui.adsabs.harvard.edu/abs/2018ApJ...862...45S} {862, 45}

\bibitem[\protect\citeauthoryear{{Stasi{\'n}ska}}{{Stasi{\'n}ska}}{2002}]{stasinska:2002}
{Stasi{\'n}ska} G.,  2002, in {Henney} W.~J.,  {Franco} J.,   {Martos} M.,
  eds,  Revista Mexicana de Astronomia y Astrofisica Conference Series Vol. 12,
  Revista Mexicana de Astronomia y Astrofisica Conference Series. pp 62--69
  (\mn@eprint {arXiv} {astro-ph/0102403}),
  \mn@doi{10.48550/arXiv.astro-ph/0102403}

\bibitem[\protect\citeauthoryear{{Stasi{\'n}ska}}{{Stasi{\'n}ska}}{2005}]{Stasinska:2005}
{Stasi{\'n}ska} G.,  2005, \mn@doi [\aap] {10.1051/0004-6361:20042216}, \href
  {https://ui.adsabs.harvard.edu/abs/2005A&A...434..507S} {434, 507}

\bibitem[\protect\citeauthoryear{{Stasi{\'n}ska} \& {Szczerba}}{{Stasi{\'n}ska}
  \& {Szczerba}}{2001}]{Stasinska:2001}
{Stasi{\'n}ska} G.,  {Szczerba} R.,  2001, \mn@doi [\aap]
  {10.1051/0004-6361:20011403}, \href
  {https://ui.adsabs.harvard.edu/abs/2001A&A...379.1024S} {379, 1024}

\bibitem[\protect\citeauthoryear{{Stasi{\'n}ska}, {Morisset},
  {Sim{\'o}n-D{\'\i}az}, {Bresolin}, {Schaerer}  \& {Brandl}}{{Stasi{\'n}ska}
  et~al.}{2013}]{stasinska:2013}
{Stasi{\'n}ska} G.,  {Morisset} C.,  {Sim{\'o}n-D{\'\i}az} S.,  {Bresolin} F.,
  {Schaerer} D.,   {Brandl} B.,  2013, \mn@doi [\aap]
  {10.1051/0004-6361/201220428}, \href
  {https://ui.adsabs.harvard.edu/abs/2013A&A...551A..82S} {551, A82}

\bibitem[\protect\citeauthoryear{{Storey}, {Sochi}  \& {Bastin}}{{Storey}
  et~al.}{2017}]{Storey:2017}
{Storey} P.~J.,  {Sochi} T.,   {Bastin} R.,  2017, \mn@doi [\mnras]
  {10.1093/mnras/stx1189}, \href
  {https://ui.adsabs.harvard.edu/abs/2017MNRAS.470..379S} {470, 379}

\bibitem[\protect\citeauthoryear{{Torres-Peimbert}, {Peimbert}  \&
  {Pena}}{{Torres-Peimbert} et~al.}{1990}]{torrespeimbertetal90}
{Torres-Peimbert} S.,  {Peimbert} M.,   {Pena} M.,  1990, A\&A, \href
  {https://ui.adsabs.harvard.edu/abs/1990A&A...233..540T} {233, 540}

\bibitem[\protect\citeauthoryear{{Truran} \& {Cameron}}{{Truran} \&
  {Cameron}}{1971}]{Truran:1971}
{Truran} J.~W.,  {Cameron} A.~G.~W.,  1971, \mn@doi [\apss]
  {10.1007/BF00649203}, \href
  {https://ui.adsabs.harvard.edu/abs/1971Ap&SS..14..179T} {14, 179}

\bibitem[\protect\citeauthoryear{{Wagoner}}{{Wagoner}}{1973}]{Wagoner:1973}
{Wagoner} R.~V.,  1973, \mn@doi [\apj] {10.1086/151873}, \href
  {https://ui.adsabs.harvard.edu/abs/1973ApJ...179..343W} {179, 343}

\bibitem[\protect\citeauthoryear{{Wesson}, {Liu}  \& {Barlow}}{{Wesson}
  et~al.}{2003}]{Wesson:2003}
{Wesson} R.,  {Liu} X.~W.,   {Barlow} M.~J.,  2003, \mn@doi [\mnras]
  {10.1046/j.1365-8711.2003.06289.x}, \href
  {https://ui.adsabs.harvard.edu/abs/2003MNRAS.340..253W} {340, 253}

\bibitem[\protect\citeauthoryear{{Wesson}, {Liu}  \& {Barlow}}{{Wesson}
  et~al.}{2005}]{Wesson:2005}
{Wesson} R.,  {Liu} X.~W.,   {Barlow} M.~J.,  2005, \mn@doi [\mnras]
  {10.1111/j.1365-2966.2005.09325.x}, \href
  {https://ui.adsabs.harvard.edu/abs/2005MNRAS.362..424W} {362, 424}

\bibitem[\protect\citeauthoryear{{Wesson}, {Jones}, {Garc{\'\i}a-Rojas},
  {Boffin}  \& {Corradi}}{{Wesson} et~al.}{2018}]{Wesson:2018}
{Wesson} R.,  {Jones} D.,  {Garc{\'\i}a-Rojas} J.,  {Boffin} H.~M.~J.,
  {Corradi} R.~L.~M.,  2018, \mn@doi [\mnras] {10.1093/mnras/sty1871}, \href
  {https://ui.adsabs.harvard.edu/abs/2018MNRAS.480.4589W} {480, 4589}

\bibitem[\protect\citeauthoryear{{Wyse}}{{Wyse}}{1942}]{wyse42}
{Wyse} A.~B.,  1942, \mn@doi [ApJ] {10.1086/144409}, \href
  {https://ui.adsabs.harvard.edu/abs/1942ApJ....95..356W} {95, 356}

\bibitem[\protect\citeauthoryear{{Yuan}, {Liu}, {P{\'e}quignot}, {Rubin},
  {Ercolano}  \& {Zhang}}{{Yuan} et~al.}{2011}]{Yuan:2011}
{Yuan} H.~B.,  {Liu} X.~W.,  {P{\'e}quignot} D.,  {Rubin} R.~H.,  {Ercolano}
  B.,   {Zhang} Y.,  2011, \mn@doi [\mnras] {10.1111/j.1365-2966.2010.17732.x},
  \href {https://ui.adsabs.harvard.edu/abs/2011MNRAS.411.1035Y} {411, 1035}

\bibitem[\protect\citeauthoryear{{Zhang}, {Ercolano}  \& {Liu}}{{Zhang}
  et~al.}{2007}]{Zhang:2007}
{Zhang} Y.,  {Ercolano} B.,   {Liu} X.~W.,  2007, \mn@doi [\aap]
  {10.1051/0004-6361:20066564}, \href
  {https://ui.adsabs.harvard.edu/abs/2007A&A...464..631Z} {464, 631}

\makeatother
\end{thebibliography}

\end{document}